\documentclass[12pt,preprint]{aastex}
\newcommand{\Cerenkov}{\v{C}erenkov}
\usepackage{rotating}

\begin{document}


\title{Search for High Energy Gamma Rays from an X-ray Selected Blazar Sample}


\author{
I.~de~la~Calle~P\'erez\altaffilmark{1},
I.H.~Bond\altaffilmark{1},
P.J.~Boyle\altaffilmark{2},
S.M.~Bradbury\altaffilmark{1},
J.H.~Buckley\altaffilmark{3},
D.A.~Carter-Lewis\altaffilmark{4},
O.~Celik\altaffilmark{14},
W.~Cui\altaffilmark{5},
C.~Dowdall\altaffilmark{6},
C.~Duke\altaffilmark{7},
A.~Falcone\altaffilmark{5},
D.J.~Fegan\altaffilmark{8},
S.J.~Fegan\altaffilmark{9}\altaffilmark{,10},
J.P.~Finley\altaffilmark{5},
L.~Fortson\altaffilmark{2},
J.A.~Gaidos\altaffilmark{5},
K.~Gibbs\altaffilmark{9},
S.~Gammell\altaffilmark{8},
J.~Hall\altaffilmark{11},
T.A.~Hall\altaffilmark{12},
A.M.~Hillas\altaffilmark{1},
J.~Holder\altaffilmark{1},
D.~Horan\altaffilmark{9},
M.~Jordan\altaffilmark{3},
M.~Kertzman\altaffilmark{13},
D.~Kieda\altaffilmark{11},
J.~Kildea\altaffilmark{8},
J.~Knapp\altaffilmark{1},
K.~Kosack\altaffilmark{3},
H.~Krawczynski\altaffilmark{3},
F.~Krennrich\altaffilmark{4},
S.~LeBohec\altaffilmark{4},
E.T.~Linton\altaffilmark{2},
J.~Lloyd-Evans\altaffilmark{1},
P.~Moriarty\altaffilmark{6},
D.~M\"uller\altaffilmark{2},
T.N.~Nagai\altaffilmark{11},
R.A.~Ong\altaffilmark{14},
M.~Page\altaffilmark{7},
R.~Pallassini\altaffilmark{1},
D.~Petry\altaffilmark{4,16},
B.~Power-Mooney\altaffilmark{8},
J.~Quinn\altaffilmark{8},
P.~Rebillot\altaffilmark{3},
P.T.~Reynolds\altaffilmark{15},
H.J.~Rose\altaffilmark{1},
M.~Schroedter\altaffilmark{9}\altaffilmark{,10},
G.H.~Sembroski\altaffilmark{5},
S.P.~Swordy\altaffilmark{2},
V.V.~Vassiliev\altaffilmark{11},
S.P.~Wakely\altaffilmark{2},
G.~Walker\altaffilmark{11},
T.C.~Weekes\altaffilmark{9} }


\altaffiltext{1}{Department of Physics, University of Leeds,
Leeds, LS2 9JT, Yorkshire, England, UK}

\altaffiltext{2}{Enrico Fermi Institute, University of Chicago,  Chicago, IL 60637}

\altaffiltext{3}{Department of Physics, Washington University, St.~Louis,
MO 63130}

\altaffiltext{4}{Department of Physics and Astronomy, Iowa State
University, Ames, IA 50011}

\altaffiltext{5}{Department of Physics, Purdue University, West
Lafayette, IN 47907}

\altaffiltext{6}{School of Science, Galway-Mayo Institute of Technology,
Galway, Ireland}

\altaffiltext{7}{Physics Department, Grinnell College, Grinnell, IA 50112}

\altaffiltext{8}{Physics Department, National University of Ireland,
Belfield, Dublin 4, Ireland}

\altaffiltext{9}{ Fred Lawrence Whipple Observatory, Harvard-Smithsonian
CfA, Amado, AZ 85645}

\altaffiltext{10}{Department of Physics, University of Arizona, Tucson, AZ 85721}

\altaffiltext{11}{High Energy Astrophysics Institute, University of Utah,
Salt Lake City, UT 84112}

\altaffiltext{12}{Department of
Physics and Astronomy, University of Arkansas at Little Rock, 
Little Rock, AR 72204}

\altaffiltext{13}{Physics Department, De Pauw University, Greencastle,
                   IN, 46135}

\altaffiltext{14}{Department of Physics, University of California, Los Angeles, CA 90095}

\altaffiltext{15}{Department of Physics, Cork Institute of Technology, Cork, Ireland}

\altaffiltext{16}{NASA/GSFC, Code 661, Greenbelt, MD 20771}


\begin{abstract}

\par Our understanding of blazars has been greatly increased in recent years
by extensive multi-wavelength observations, particularly in the radio, X-ray
and gamma-ray regions. Over the past decade the Whipple 10\,m telescope has
contributed to this with the detection of 5 BL Lacertae objects at very high
gamma-ray energies. The combination of multi-wavelength data has shown that
blazars follow a well-defined sequence in terms of their broadband spectral
properties. Together with providing constraints on emission models, this
information has yielded a means by which potential sources of TeV emission may
be identified and predictions made as to their possible gamma-ray flux. We
have used the Whipple telescope to search for TeV gamma-ray emission from
eight objects selected from a list of such candidates. No evidence has been
found for VHE emission from the objects in our sample, and upper limits have
been derived for the mean gamma-ray flux above 390~GeV. These flux upper
limits are compared with the model predictions and the implications of our
results for future observations are discussed.

\end{abstract}




\section{Introduction}\label{introduction}

\par According to the unified scheme of active galactic nuclei (AGN) a blazar
is considered to be any radio loud AGN which displays highly variable, beamed,
non-thermal emission from radio to gamma-ray wavelengths \citep{Urry95}. The
non-thermal emission is believed to originate from a population of electrons
moving with relativistic velocity along a plasma jet oriented with a small
angle to the line of sight \citep{Sikora94}. In a
$\nu \mathrm{F}(\nu)$ plot, the spectral energy distribution (SED) of blazars shows two
broad peaks, one at energies ranging from the infrared to the X-ray band and
the other in the gamma-ray band. Two approaches have been taken to
characterize the blazar SED: a phenomenological approach
\citep{Fossati98}, where the bolometric source luminosity largely governs the
blazar SED, and more model-dependent approaches. Several models have been put
forward in an attempt to explain the blazar SED, the most frequently cited of
which is the Synchrotron-Self-Compton (SSC) model. In this model, the origin
of the first peak is synchrotron emission from relativistic electrons which are
then upscattered by the inverse Compton (IC) process to form the second peak
in the gamma-ray band \citep{Bloom96}. In External Compton (EC) models,
the synchrotron emission may be accompanied by ambient soft photons of
different origin as a target for the IC process \citep{ Dermer93, Sikora94a,
Ghisellini96b}.

\par BL Lacertae (BL Lac) objects are a blazar sub-class, characterised by
their lack of strong emission or absorption lines, in which the second peak of
the SED can extend up to very high (GeV-TeV) energies. They are therefore
preferred targets of ground-based observations with atmospheric {\Cerenkov}
telescopes. Using these observations to determine the position of the second
peak of the SED allows us to constrain associated parameters of emission
models, most importantly, the intensity of the magnetic field and the maximum
energy to which the particles responsible for the emission are accelerated
\citep{Buckley99}.

\par TeV gamma rays from extragalactic sources suffer absorption by
interaction with photons from the IR background radiation \citep{Gould66,
Primack99, deJagger02}. A large TeV BL Lac sample containing objects at a
range of redshifts would provide the means to distinguish between intrinsic
spectral features and the effects of absorption by the infrared
background. However, at present, the list of known TeV BL Lacs is very small
and needs to be expanded. The limited field of view of {\Cerenkov} telescopes
and their low duty cycle forces us to select, {\it a priori}, the most
promising target objects from a catalogue of candidates for study. In the past
the Whipple Collaboration has targeted the nearest northern hemisphere BL Lacs
and this approach resulted in several positive detections: Mrk~421 ($z=0.030$)
\citep{Punch92}, Mrk~501 ($z=0.034$) \citep{Quinn96} and 1ES2344+514
($z=0.044$) \citep{Catanese98}. In addition, observations of BL Lacs with the
peak of the first component of their SED extending far into the X-ray band
($>0.1$~keV) led to the detection of the most distant TeV source 1ES1426+428
($z=0.129$) \citep{Horan02}. Most recently, 1ES1959+650 \citep{Holder02} was
detected at TeV energies in an intense flaring state, using the Whipple
telescope. This object is also a close BL Lac ($z=0.048$) whose status as a
TeV source was confirmed following marginal detection in 1998 at TeV energies
by the Utah Seven Telescope Array collaboration \citep{Nishiyama99}.

\par In this paper, we report on candidate TeV sources selected following the
work of \citet{Costamante01}. Their BL Lac catalogue is the first to provide
estimates of TeV fluxes based on detailed model predictions and includes
successful predictions of the TeV flux of 1ES1959+650 in its flaring state
\citep{Aharonian03}. The catalogue consists of objects bright in {\it both}
the X-ray and radio bands; i.e. objects with {\it both} electrons energetic
enough to upscatter seed photons to TeV energies {\it and} a high density of
seed photons to be upscattered. The radio and X-ray fluxes are plotted in
Figure \ref{fxfrk_zoom} and, as can be seen, the sample includes established
TeV sources (including PKS2155-304 ($z=0.116$) \citep{Chadwick99}). The
gamma-ray flux at TeV energies is estimated by applying a homogeneous one-zone
SSC model \citep{Ghisellini02} and by using the phenomenological
parameterisation of the blazar SED developed by \citet{Fossati98} and adapted
by \citet{Costamante01}. The latter approach by \citet{Costamante01} was
motivated by a desire to better describe the SEDs of low power blazars. These
flux estimates allow us to assess the detectability of this sample of BL Lacs
using present, or future, more sensitive, atmospheric {\Cerenkov}
telescopes. However, the dramatic variability of these sources significantly
adds to the uncertainty of detecting a source at any given time.

\par We observed eight objects selected from this sample (Section
\ref{Selection}). In Sections \ref{Observations} and \ref{Analysis} we present
a summary of the TeV data and discuss a new analysis method of background
estimation suited to our observation strategy. The upper limits derived from
our observations are summarised in Section \ref{UpperLimits} and discussed
within the framework of the popular emission models. Our observations are
compared with RXTE All Sky Monitor (ASM) X-ray data where available (Section
\ref{XGCurves}). In Section \ref{Discussion} we discuss implications of non
detections of these candidates for future observations.

\section{Selection of AGNs for Whipple Observations}\label{Selection}

\par We have chosen from the BL Lac candidate list developed by
\citet{Costamante01}, the most suitable candidates for TeV observations. We
preferentially selected objects which culminate at an angle to the zenith
$\Theta < 30^\circ$ and objects with redshifts smaller than 0.2, where opacity
due to pair production is not so extreme as to prevent detection. Based on the
prediction of \citet{Fossati98}, we required that the flux above 0.3~TeV
should be greater than $\sim~10\%$ of the Crab Nebula flux so as to keep
observing time below a reasonable value of $\sim~50$ hours per source. In some
cases the selection of objects was biased by the past record of X-ray
activity, i.e., episodes of flaring activity.

\par After the selection criteria were applied, 6 objects were identified as
the best target candidates for TeV observations by the Whipple 10\,m
telescope during the 2001-2002 observing season (circled sources in Figure
\ref{fxfrk_zoom}): 1ES0033+595, RGBJ0214+517, 1ES0229+200, 1ES0806+524,
RGBJ1117+202 and RGBJ1725+118. Although at a higher redshift, two extra
objects, 1ES0120+340 and 1ES1553+113, were included in this work since both
belong to the list of TeV candidates and were observed in the same season by
the Whipple telescope. The selection of these two objects was based upon
their extreme nature \citep{Ghisellini99}, both having many similarities in
their broadband properties to 1ES1426+428. Table \ref{predicted_flux_levels}
lists the selected objects along with the predicted flux values.

\section{Observations}\label{Observations}

\par VHE observations reported here have been made with the Whipple
Observatory 10\,m gamma-ray Telescope \citep{Cawley90} on Mt. Hopkins,
Arizona. {\Cerenkov} radiation produced by gamma-ray and cosmic-ray induced
atmospheric showers is recorded by a high resolution camera located on the
focal plane. The recent camera \citep{Finley01} is equipped with 379
$0.12^\circ$ photomultiplier tubes (PMTs), giving a total field of view (FOV)
of $2.6^\circ$ diameter. This inner part of the camera is surrounded by three
circular rings of 111 $0.25^\circ$ PMTs which extend the FOV to $4^\circ$
diameter. The telescope uses a hardware pattern recognition trigger which
suppresses accidental triggers due to the night sky background light
\citep{Bradbury}. Only the inner 331 PMTs participate in the telescope
trigger. Although the signals from the outer 111 PMTs are recorded by the
electronics, they are not involved in the analysis process used here. Table
\ref{TelescopePerformance} shows the telescope's performance based on of Crab
Nebula observations, the standard candle for TeV astronomy, which were taken
over the same period of time as the observations reported here. The telescope
energy threshold (defined as the energy at which the response to a Crab-like
source peaks) was 390 $\pm$ 80$_{sys}$~GeV during the same period of time.

\par Observations were carried out between 2001 October and 2002 July, with
total observing times on individual sources ranging from a few hours to about
20 hours and restricted to zenith angles less than 30$^\circ$. For comparison
purposes we also summarise observations on the detection at TeV energies of
1ES1426+428 and 1ES1959+650. The observations of 1ES1426+428 reported in
\citet{Horan02} correspond to the period of time between 2001 February and
2001 June when the object was observed most extensively and a detection was
claimed above a 5$\sigma$ level by the Whipple Collaboration. 1ES1959+650 data
\citep{Holder02} are taken from observations of the source during a flaring
state between 2002 May and 2002 July. All observations are summarised in Table
\ref{coord_table}.

\par The database consists of ON-source observations, typically each of 28
minutes duration. A subset of observations are accompanied by control
OFF-source observations which target a region of the sky free from known
gamma-ray sources. The OFF-source observation is taken directly before or
after the ON-source observation on a region offset 30 minutes in right
ascension from the ON-source region, thereby following the same
elevation-azimuth path in the sky as the candidate source.

\par Predicted flux levels based on the parameterised SED of \citet{Fossati98}
(L. Costamante 2001, private communication) have been used to estimate how
much ON-source observing time would be required to achieve a 5$\sigma$
detection for each source. These flux estimates give us the most optimistic
time exposure required to achieve a significant detection. Table
\ref{predicted_flux_levels} shows the predicted flux values along with the
required ON-source observing time. Required observing times have been obtained
based on the telescope sensitivity to a Crab-like source (and assuming
background rates similar to those of the Crab OFF-source region).


\par The stability of cosmic-ray trigger rates and throughput factor, a
measurement of the change in telescope efficiency according to the procedure
described in \citet{Bohec02}, are monitored every observing night and have been
used in this work to select individual data runs in order to provide good
quality data sets.

\section{Analysis Technique}\label{Analysis}

\par The data have been subjected to the standard image processing analysis
where recorded {\Cerenkov} images are parameterised using a moment analysis
\citep{Hillas85}. A set of image parameter cuts (Supercuts 2000), optimised on
Crab Nebula data, was applied in order to identify candidate gamma-ray events
based on the image shape (\textit{width} and \textit{length}), location
(\textit{distance}) and orientation (\textit{alpha}) \citep{Reynolds93}. The
image parameter domain used in this work in order to select gamma-ray
candidate events, is summarised in Table \ref{SC2000}.


\par The aim of any analysis method is to estimate the mean number of photons
coming from the source direction, which requires a reliable estimate of the
number of background cosmic-ray events. Ideally, an equal exposure of
ON-source observations and control OFF-source observations is used; however in
an attempt to increase the total ON-source exposure times, many runs are taken
with no accompanying control OFF-source observations. Prior to this work, the
background for these observations was directly estimated from the ON-source
observations themselves, using events with orientations \textit{alpha} such
that they are not from the direction of the source \citep{Catanese98}. To avoid
possible systematic effects introduced by this method when estimating the
statistical significance of an excess, an alternative method of determining the
number of background events has been considered. The concern is that in order to
reach the predicted flux levels for the sources considered in this work, long
exposure times are required to achieve a significant detection. We therefore
need to ensure that systematic errors are kept to a minimum.

\par The method involves selecting \textit{any} OFF-source observation,
recorded at similar elevations and preferably on the same night as the
ON-source observation of interest. This `matching' procedure is carried out
qualitatively by comparing five different factors: date, throughput factor,
mean elevation, mean sky noise and number of pixels turned off during data
taking (to avoid bright stars in the field of view). The first two are chosen
to ensure that the telescope performance and weather conditions are the same
for ON-source and OFF-source observations. The other three parameters are
chosen to ensure as much as possible that the shape of the image parameter
distributions is similar within statistical fluctuations.

\par Once ON/OFF matched pairs are obtained, standard ON/OFF analysis is
performed, including software padding \citep{Cawley93} to compensate for
differences in sky brightness between the ON-source observations and the
corresponding control OFF-source observations. After matching and padding, the
distribution of the parameter \textit{alpha} for the ON-source and OFF-source
observations is obtained. Despite the care taken to match ON-source and
OFF-source observations, a remaining difference in the total number of events
after image cuts, due largely to small differences in the zenith angles of the
matched pairs, has still to be corrected for. This is done by introducing a
scaling factor (SCF) which scales both \textit{alpha} distributions,
corresponding to the ON-source and OFF-source observations, using the total
number of events in the alpha region from $30^\circ$ to $90^\circ$. Table
\ref{tev_observations} shows the SCF for the sources considered in this work
(in brackets in the column headed {\it Events}).  Where genuine ON/OFF pairs
are available, these are added to the matched pairs before scaling the
\textit{alpha} distributions.

\par When no significant excess of a signal over the background is found, it
is possible to set an upper limit to the mean gamma-ray flux expected from the
source direction. Calculations of flux upper limits depend on the number of
background events and the number of events coming from the source
direction. In this work, flux upper limits are calculated in several steps, in
a similar way to the approach of \citet{Aharonian00a}. Firstly, the method of
\citet{Helene} is applied to set an upper limit to the mean number of
gamma-ray events expected from the source direction. When applying this method
we required that the number of background events is always smaller than the
number of events coming from the source direction, i.e. that the mean number
of source counts is always positive. If the opposite is true a conservative
approach is taken and the number of events coming from the source direction is
made equal to the number of background events. Secondly, the upper limit in
source counts is converted into a fraction of the Crab Nebula flux by
comparison with contemporaneous Crab Nebula observations (UL$_\mathrm{c}$). The
conversion into absolute flux units is straightforward using the Crab Nebula
flux \citep{Hillas98} above the energy threshold, E$_{\mathrm{th}}$, of the
observations (F$_{\mathrm{Crab}} (>0.39$~TeV) = 8.30 $\times$
$10^{-11}$~erg~cm$^{-2}$~s$^{-1}$). A more general equation can be used (see
eq. [\ref{Upper_Limit_eq}]) to obtain the upper limit using the Crab Nebula flux
above any given energy, E:

\begin{center}
\begin{equation}
UL(>E_{th}) = UL_\mathrm{c} F_{\mathrm{Crab}}(>E)
[\frac{E_{\mathrm{th}}}{E}]^{-\alpha + 1} \quad \quad \quad \mathrm{cm}^{-2} \mathrm{s}^{-1}
\label{Upper_Limit_eq}
\end{equation}
\end{center}

\par This method assumes that the Crab Nebula flux is stable over time
\citep{Hillas98} and that the differential spectrum ($\alpha$) of the putative
source is similar to that of the Crab Nebula ($\propto E^{-\alpha}$ where
$\alpha$=2.49). The only uncertainty to be taken into consideration on the
flux upper limits is that introduced by the normalization of the Crab Nebula flux,
i.e. the uncertainty in the Crab Nebula photon flux. Above the telescope
energy threshold (390~GeV) the uncertainty is $25\%$ and above 1~TeV, $10\%$
\citep{Catanese98}. The amount of IR absorption plays an important role in
modelling the blazar SED at high energies (the emission models should account
for the intrinsic spectra and not the IR absorbed ones) so estimates have been
obtained for the amount of IR absorption expected in each case according to
the source redshift and assuming again a Crab-like source spectrum.


\section{Results}\label{Results}

\subsection{Flux Upper Limits}\label{UpperLimits}

\par Table \ref{tev_observations} summarises the results of the complete set
of observations on the 8 selected sources. As this class of source is known to
be highly variable, light curves (source count rates vs. time) have been used
to search for episodes of emission on three different timescales: 28 minute,
daily, and monthly. No evidence for variable VHE emission has been seen over
any of these time scales. Flux upper limits are given at a $97\%$ ($\sim
2\sigma$) confidence level (c.l.) in Crab and absolute flux units above the
energy threshold of 390~GeV. Upper limits are also reported above 1~TeV by
reducing the flux at 390~GeV by 75$\%$, which assumes again a Crab-like
spectrum.

\par Table \ref{tev_comparisson} shows the flux upper limits derived in this
work as compared to the two predicted fluxes given in
\citet{Costamante01}. The estimated percentage of absorption of the gamma-ray
photon flux due to the IR background is also given. For two of the objects in
our sample, RGBJ0214+517 and RGBJ1725+118, the flux upper limit, IR absorption
corrected, is below the value predicted by the adapted version of the Fossati
parameterisation above 0.3~TeV. According to this parameterisation, the
predicted fluxes are derived from SEDs that represent the average state of a
source in a given radio luminosity range. The SEDs were derived including also
those of the known TeV blazars; therefore it is considered that the derived
flux estimates from these SEDs are more representative of a high emission
state. On the contrary, the SSC model considered \citep{Ghisellini02} is
designed to fit only the synchrotron spectra of the sources, more
representative of a quiescent state. The inverse Compton component is then
inferred from the synchrotron one. Upper limits derived in this work do not
contradict the SSC-predicted TeV emission above 0.3~TeV since in all cases our
flux upper limits are well above the predicted value. For 1ES0229+200 there is
a flux prediction above 0.3~TeV, of 0.02 Crab units (c.u.), by
\citet{Stecker96}. Our flux upper limit on this source is not in contradiction
with this.

\par The flux upper limits derived in this work have been compared with those
previously available from the CAT \citep{Piron} and HEGRA \citep{ICRC27,
Aharonian00a} experiments where available (Table \ref{tev_comparisson}). A
direct comparison of flux upper limits derived by the different experiments
shows that they are of the same order of magnitude. For the sources listed in
Table \ref{tev_comparisson}, flux upper limits derived by different experiments
cannot be used to invalidate each other, since observations are not
contemporaneous.

\par
\begin{sidewaysfigure}
\plotone{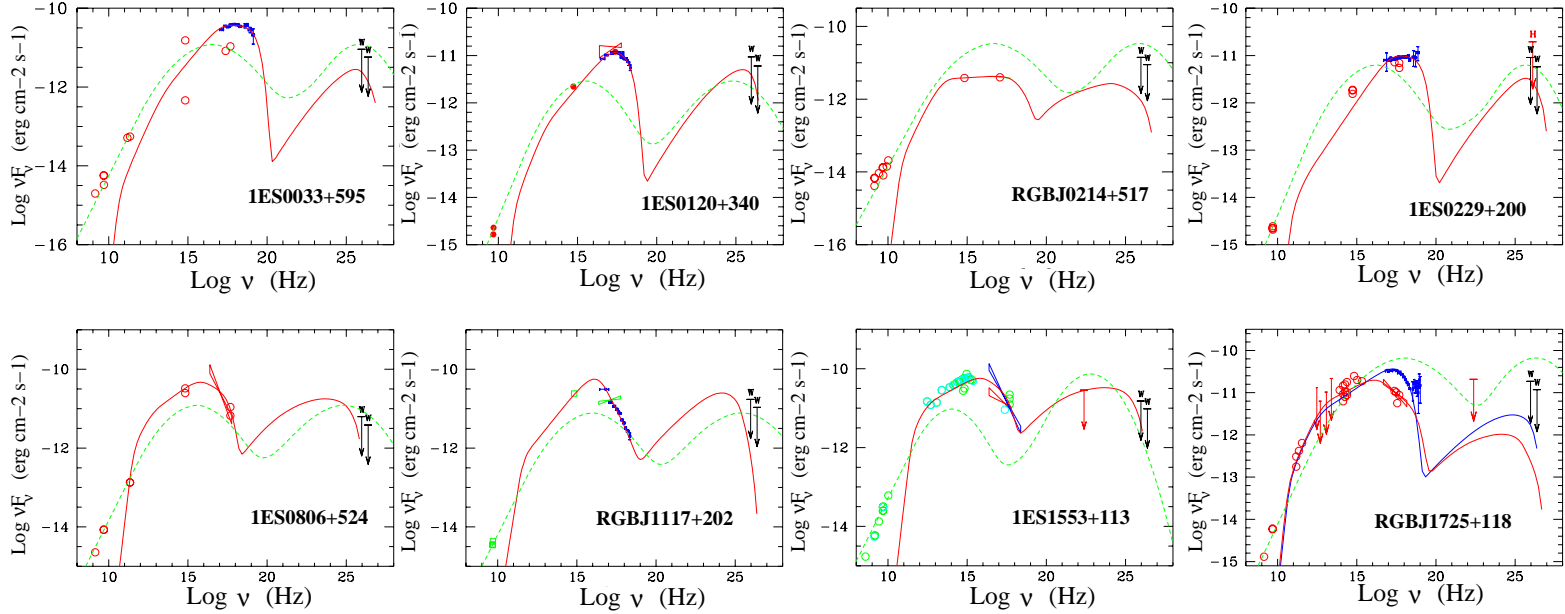}
\figcaption[SSC model of the 8 X-ray selected BL Lacs] {SSC model (solid line)
and phenomenological parameterisation of \cite{Fossati98} as modified by
\cite{Costamante01} (dashed line). From top left to bottom right, 1ES0033+595,
1ES0120+340, RGBJ0214+517, 1ES0229+200, 1ES0806+524, RGBJ1117+202, 1ES1553+113
and RGBJ1725+118. Flux upper limits obtained in this work are represented by
arrows labelled as {\it W} at two energies, 390~GeV and 1~TeV. HEGRA flux upper
limit for 1ES0229+200 has been labeled as {\it H}. (Figures
courtesy of L. Costamante).\label{SSC_Models}}
\end{sidewaysfigure}

 Figure \ref{SSC_Models} shows the SED of the 8 objects as described by an
SSC model (solid lines) and by the adapted phenomenological parameterisation
of Fossati (dashed lines) \citep{Costamante01}. Flux upper limits obtained in
this work are represented by arrows labelled as {\it W} at two energies,
390~GeV and 1~TeV.

\par We note additionally that, because of the time variability of high energy
emission from these objects, the upper limits reported in this work are only
valid for the period of time during which observations were carried out.

\subsection{ASM X-ray Fluxes}\label{XGCurves}

\par The X-ray flux of the sources in our sample can provide an indication of
whether enhanced TeV emission was expected over the period of the gamma-ray
observations. According to the SSC model, if the same population of electrons
is responsible for the two peaks observed in the blazar SED, an increase in
the synchrotron photon flux would lead to a corresponding increase in the IC
photon flux. In BL Lac objects the behaviour of the most energetic electrons
is well monitored in the X-ray band (dominated by the synchrotron emission),
and in the GeV-TeV band (dominated by IC emission). In fact, strong
correlation has been observed between the fluxes in the X-ray and GeV-TeV band
which match the predictions of the IC models (e.g. observations of Mrk501 in
1997 \citep{Catanese97b,Pian98,Krawczynski00}, observations of Mrk421 in 1995
\citep{Buckley96} and 2001 \citep{Holder01}).

\par Data from the All Sky Monitor (ASM) on board the Rossi X-Ray Timing
Explorer (RXTE) were available for 5 of the 8 sources in our sample, covering
the period from 1997 January 1 to 2002 July 31. We calculated the mean X-ray
flux over the exact period of time in which TeV gamma-ray observations
reported here were taken, the average over the years 1997 through 2002 and the
average on a yearly basis (see Table \ref{xray_meanfluxes_tab}). This mean
flux over long periods of time is not very representative for these types of
objects which show very strong variability in their emission states over a
range of timescales, from hours to years; however, from the comparison of
these mean fluxes we conclude that the mean X-ray flux over the period of time
in which TeV gamma-ray observations were taken does not increase significantly
from the mean over all years for any of the sources. Only 1ES0806+524 shows a
marginally higher than average X-ray flux during the year 2002 (in which
gamma-ray observations were taken), 2.7$\sigma$ above the average X-ray flux
over all years.

\par As a comparison, we have proceeded in the same way using ASM data
corresponding to the known TeV sources Mrk421, Mrk501, 1ES1426+428 and
1ES1959+650 and present the results in Table \ref{xray_meanfluxes_tab}. In the
cases of 1ES1426+428 and 1ES1959+650 the mean X-ray flux during the period of
time in which the TeV detection observations were taken is significantly
higher than the mean over all years. Mrk421 and Mrk501 show a higher than
average X-ray flux over the years 2001 and 1997 respectively; that is, in
those two years when both objects showed periods of the highest gamma-ray
activity observed (flux levels of up to 13 times that of the Crab Nebula flux
in the case of Mrk421 \citep{Krennrich01} and up to 4 times that of the Crab
Nebula flux in the case of Mrk501 \citep{Catanese97b}).  However, there have
been occasions when rapid X-ray or TeV flares have been seen without a
corresponding flux variation in the other energy band (e.g. Mrk501
\citep{Catanese00} and 1ES1959+650 \citep{Krawczynski03}).

\par From this simple study it is inferred, if we assume that the properties
described above are representative of the general behaviour of BL Lac objects,
that it is possible to use the mean X-ray flux, at least over long time
scales, as an indicator of source activity, which in turn could result in
enhanced emission of TeV gamma rays. Our sources show no evidence of a
long-lasting period of high activity, as indicated by the average X-ray flux,
over the period of time in which our TeV observations took place. Even for the
well-established TeV sources, Mrk421 and Mrk501, there are periods of time in
which the TeV fluxes fall below the sensitivity of current {\Cerenkov}
telescopes and are not detected, corresponding with periods of low X-ray
activity \citep{Piron01}. Also, it is inferred that the level of the
mean X-ray flux can be used as a potential indicator of gamma-ray activity
and might trigger follow-up gamma-ray observations.

\section{Discussion}\label{Discussion}

\par In this work we have observed a set of BL Lac objects selected from a
sample of promising candidates for TeV emission, developed by
\citet{Costamante01}. This sample is comprised of objects that show high radio
and X-ray fluxes since in order to produce a strong TeV signal by the IC
process, a large number of energetic electrons and seed photons is
required. These authors have also provided estimates of the TeV fluxes, according
to a detailed SSC emission model, which can be used both to establish the
detectability of this sample of objects and to test the model itself. Our TeV
observations have resulted in a set of flux upper limits, above an energy of
390~GeV, which do not conflict with the predicted values according to current
SSC models. We investigate here the possible reasons for these non-detections.

\par The first possibility is that the TeV fluxes are far below the
sensitivity of current {\Cerenkov} telescopes. The SSC model used to derive
the flux estimates at TeV energies predicts TeV fluxes more representative of
a quiescent state of the sources, and the estimated times required to
guarantee a 5$\sigma$ detection are all above 500 hours (see Table
\ref{predicted_flux_levels}). Considering that the available observing time
with the Whipple telescope amounts to 700 - 800 hours per season, spending 500
hours of an observing season on a single source is not feasible. The current
sensitivity of the Whipple telescope does not allow us to test the
detectability of this sample of objects, i.e. our TeV observations do not
allow a test of this SSC model. However, the next generation of {\Cerenkov}
telescopes, such as VERITAS-4 \citep{VERITAS}, with a sensitivity above
300~GeV of $8 \times 10^{-13}$~erg~cm$^{-2}$~s$^{-1}$ (5~mCrab) on a Crab-like
source spectrum (50 hours for a 5$\sigma$ detection), would be able to reach
the predicted flux levels in just a few hours of observations.

\par The second reason for our non-detections lies in the fact that the
sources were not in a high flux emission state. Our simple study in Section
\ref{XGCurves} suggests that, at least for long-lasting flares, there is
a correlation between X-ray and gamma-ray fluxes for the well established TeV
blazars. We have shown that, for some of the objects in our sample, the ASM
X-ray fluxes show no significant increase over the period of time during which
the TeV observations reported here took place. Therefore, we may not
have detected these objects at TeV energies because they were in a low flux
emission state during this period of time.

\par Another possible reason for our non-detections could be that the sample
used to select targets for TeV observations does not provide a reliable list
of candidate BL Lacs for TeV emission. However, this explanation is weakened
by the recent detections of 1ES1426+428 and 1ES1959+650 at TeV energies. These
two objects belong to the candidate list developed by \citet{Costamante01} and
have been detected at different emission states. The detection of 1ES1426+650
by the Whipple Collaboration came as a result of a long observing campaign
during the year 2001 and since then there has been no strong evidence for TeV
emission. On the contrary, 1ES1959+650 was detected on a nightly basis in a
flaring state at a flux level of up to 5~Crab in May through July 2002. Since
then no significant detection has been reported for that source. Due to the
extreme variability that this type of object exhibits, our non-detections
should not discourage future observations of objects selected from this
sample.

\par The detection of 1ES1959+650 at TeV energies could be seen as an example
to encourage future observations of this sample of objects. A flux upper limit
of $F_{E>0.35~\mathrm{TeV}}^{99.9\%~\mathrm{c.l.}}$ = 0.13 c.u. was derived by
\citet{Catanese97a} from observations in 1996, which is far above the flux
level of 3~mCrab, above 0.3~TeV, predicted by the SSC model. With the Whipple
telescope sensitivity, at a flux level of 3~mCrab, more than 26,000 hours of
observations would have been required to achieve the detection of a gamma-ray
signal at a 5$\sigma$ level. However, the HEGRA Collaboration reported the
detection of this source in a more quiescent state from $\sim$ 95 hours
observations during the 2000 and 2001 with the HEGRA system of telescopes
\citep{Aharonian03}. The VHE flux reported for this source during that period
of time was at a level of $5.3\%$ that of the Crab Nebula flux. At this level
250 hours of observations with the Whipple telescope would be required for a
significant detection.

\par
\begin{figure}
\plotone{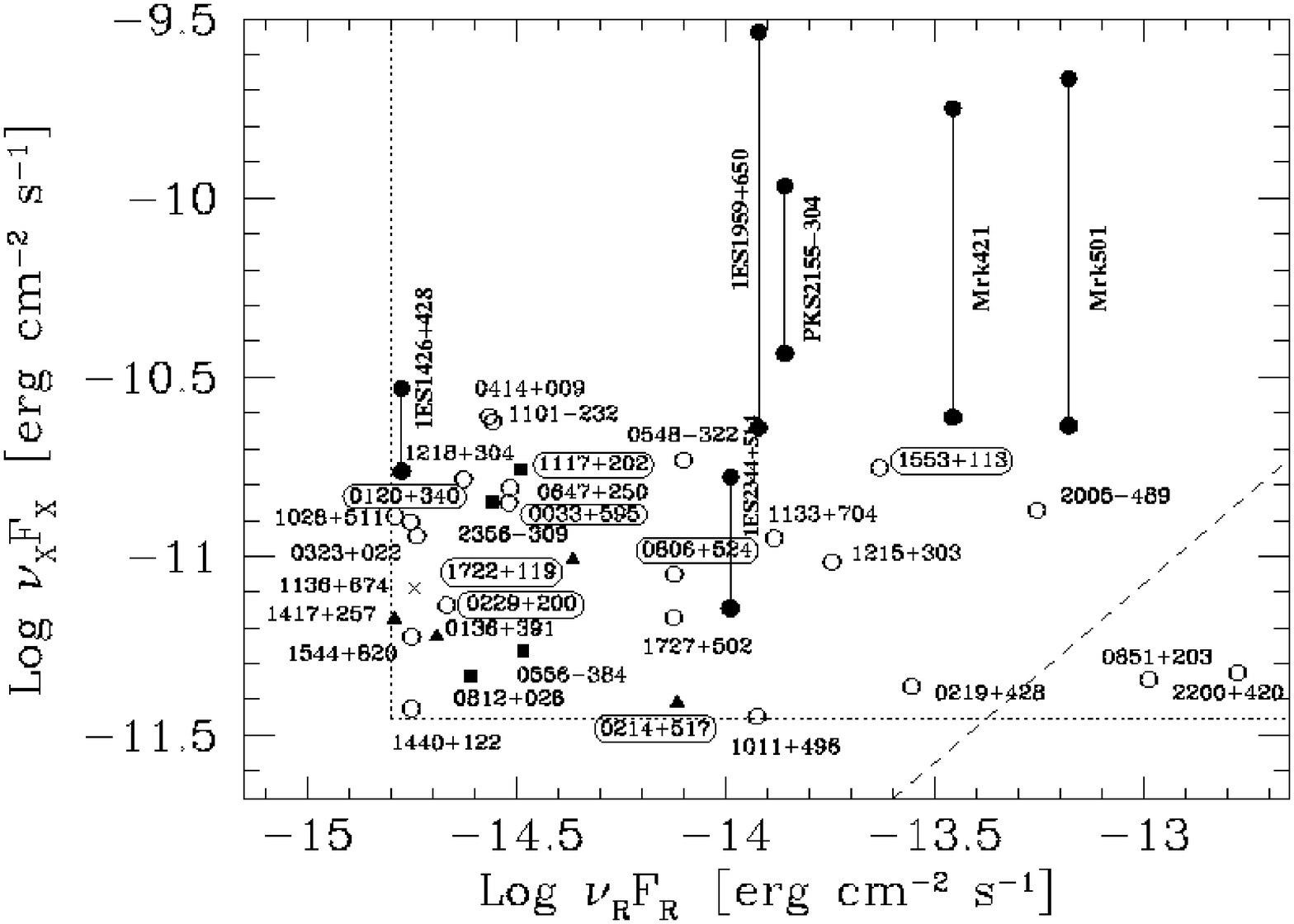} \figcaption{BL Lac objects in a $\nu \mathrm{F}(\nu)$
plane. F$_\mathrm{R}$ and F$_\mathrm{X}$ represent the radio and X-ray flux at
$\nu_\mathrm{R}$=5~GHz and $\nu_\mathrm{X}$=1~keV respectively. Different
symbols indicate different samples (see \cite{Costamante01} for
details). Encircled source names correspond to the objects considered in this
work. Black circles correspond with two flux states, quiescent and flaring, of
the known TeV sources (figure courtesy of L. Costamante).
\label{fxfrk_zoom}}
\end{figure}

 Further work is in progress to extend the list of objects bright in both
the X-ray and the radio band without pre-selection for source type. The ROSAT
Bright Source Catalogue \citep{Voges99} has been used to produce a list of
objects that would fall in the region delimited by Figure
\ref{fxfrk_zoom}. 3C~120 ($z=0.0334$) was selected from this list as one of
the brightest objects in both bands. 3C~120 is a Seyfert I galaxy with the
angle between the jet and the line of sight $<$ 20$^\circ$. Whipple
observations (7.5~hours) during the 2002-2003 observing season resulted in a
flux upper limit of $1.74 \times 10^{-11}$~cm$^{-2}$~s$^{-1}$ (20$\%$ that of
the Crab Nebula flux). Also very recently, the work by \citet{Padovani03} has
identified a new population of flat-spectrum radio quasars (FSRQ) with SEDs
resembling those of high-energy-peaked BL Lacs which are potential candidates
for TeV emission. Observations of these new types of candidate, as well as
continuing observations of the sources discussed in this paper, are in
progress now with the Whipple telescope and will continue with the telescope
array system VERITAS-4 which is currently under construction.

\acknowledgments The VERITAS Collaboration is supported by the U.S. Dept. of
Energy, N.S.F., the Smithsonian Institution, PPARC (U.K.) and Enterprise
Ireland. We acknowledge the technical assistance of E. Roache and
J. Melnick. We thank L. Costamante for his help and comments as well as the
assistance and discussions of J. Holder throughout this work.





\clearpage

\clearpage

\clearpage

\begin{deluxetable}{lcccccc}
\tabletypesize{\scriptsize} 
\tablecaption{Predicted TeV flux values for the BL Lac objects considered in
this work in absolute flux (f.u.) and Crab units (c.u.). Two estimates are
given for each source, one obtained from the parameterisation of the SED
adapted from \cite{Fossati98} (FOSS) the other from the SSC model by
\cite{Costamante01}. Conversion to Crab units has been done using a Crab flux
above 0.3~TeV of F$_{\mathrm{Crab}}(>$0.3~TeV) = 12.27 $\times$
$10^{-11}$~erg~cm$^{-2}$~s$^{-1}$. The observing time corresponds to the time 
needed for a 5$\sigma$ detection given a sensitivity of $5.74/\sqrt{t(h)}$
(Table \ref{TelescopePerformance}). Flux values in bold are the ones used to
estimate the observing time. The actual time spent ON-source is also given as a
reference. (Flux units, f.u., $10^{-11}$~erg~cm$^{-2}$~s$^{-1}$). 
\label{predicted_flux_levels}}
\tablewidth{0pt}
\tablehead{
            & \multicolumn{4}{c}{Predicted Flux} & \multicolumn{2}{c}{Observing Time} \\ 

            & \multicolumn{2}{c}{F($>$0.3~TeV)} & \multicolumn{2}{c}{F($>$1~TeV)}& \\

            & FOSS/SSC & FOSS/SSC & FOSS/SSC & FOSS/SSC & Required & Exposure \\

Source      & (f.u.)   & (c.u.)   & (f.u.)   & (c.u.)   & (h)      & (h) \\
}
\startdata
\multicolumn{1}{l}{\bf 1ES0033+595}  & 2.04/0.25   & {\bf 0.166/0.021}  &
0.48/0.04   & \multicolumn{1}{c}{0.229/0.019} & 23/1482   &  12.02\\

\multicolumn{1}{l}{\bf 1ES0120+340}  & 0.28/0.30   & {\bf 0.024/0.025}  &
0.06/.....    & \multicolumn{1}{c}{0.029/.....}   & 1135/1047 &   5.05\\

\multicolumn{1}{l}{\bf RGBJ0214+517} & 5.93/0.07   & {\bf 0.483/0.006}  &
1.43/6.2E-3 & \multicolumn{1}{c}{0.681/0.003} & 2.8/18107 &   6.05\\

\multicolumn{1}{l}{\bf 1ES0229+200}  & 0.96/0.31   & {\bf 0.078/0.026}  &
0.21/4.0E-3 & \multicolumn{1}{c}{0.100/0.002} & 101/968   &  14.69\\

\multicolumn{1}{l}{\bf 1ES0806+524}  & 1.36/.....    & {\bf 0.111/.....}    &
0.27/.....    & \multicolumn{1}{c}{0.129/.....}   & 50/.....    &  18.70\\

\multicolumn{1}{l}{\bf RGBJ1117+202} & 1.17/0.10   & {\bf 0.095/0.008}  &
0.28/.....    & \multicolumn{1}{c}{0.133/.....}   &  68/10189 &   3.26\\

\multicolumn{1}{l}{\bf 1ES1553+113}  & 0.20/0.42   & {\bf 0.016/0.035}  &
0.02/.....    & \multicolumn{1}{c}{0.010/.....}   & 2260/535  &   2.82\\ 

\multicolumn{1}{l}{\bf RGBJ1725+118} & 12.80/0.015 & {\bf 1.043/0.001}  &
3.52/1.0E-3 & \multicolumn{1}{c}{1.676/.....}   &0.67/651240&   2.33\\
\enddata
\end{deluxetable}

\clearpage

\begin{deluxetable}{lccccc}
\tablecaption{2000 - 2001 and 2001 - 2002 Whipple telescope sensitivity obtained from 
Crab Nebula observations. $\langle \Theta \rangle$ refers to the mean zenith
angle of the observations. \label{TelescopePerformance}}
\tablewidth{0pt}
\tablehead{
\multicolumn{1}{c}{Observing}  & $\langle \Theta \rangle$ & Exposure & Excess Events & Bkg$^{(*)}$ Events & Sensitivity \\ 
\multicolumn{1}{c}{Season}     &     $(^\circ)$           &  (h)     & ($\gamma$/min)  & (counts/min) & ($\sigma$/$\sqrt{(h)})$   \\
}
\startdata
2000 - 2001 & 19.7 & 8.3   & 3.36 $\pm$ 0.20 &  8.66 $\pm$ 0.13 & 5.74 \\
\tableline  	           
2001 - 2002 & 18.7 & 23.7  & 2.75 $\pm$ 0.10 &  5.88 $\pm$ 0.06 & 5.57 \\
\enddata
\tablenotetext{(*)}{Background}
\end{deluxetable}

\clearpage

\begin{deluxetable}{lccccccc}
\tabletypesize{\scriptsize}
\tablecaption{Summary of observations carried out during the 2000-2001 and 2001-2002
observing seasons$^{(*)}$ with the Whipple 10\,m telescope of 10 X-ray selected
blazars. $\langle \Theta \rangle$ refers to the mean zenith angle of the
observations. \label{coord_table}}
\tablewidth{0pt}
\tablehead{
& \multicolumn{2}{c}{Equatorial Coordinates} & & \multicolumn{2}{c}{Observing Period}
& \multicolumn{2}{c}{Whipple Observatory} \\
       & r.a. &  dec.  & Redshift & Season & MJD & Max Ele
& $\langle \Theta \rangle$ \\
Source & (J2000) & (J2000) & & & & ($^\circ$) & ($^\circ$) \\
}
\startdata
\multicolumn{1}{l}{\bf 1ES0033+595}  & $00^h35^m52\fs63$ & $+59^\circ50'04\farcs60$ & 
0.086$^{(a)}$ & Oct2001-Jan2002 & 52193-52283 & 62.1  & 29.0 \\

\multicolumn{1}{l}{\bf 1ES0120+340}  & $01^h23^m08\fs55$ & $+34^\circ20'47\farcs50$ & 
0.272 & Oct2001-Nov2001 & 52195-52234 & 87.6  & 8.0  \\

\multicolumn{1}{l}{\bf RGBJ0214+517} & $02^h14^m17\fs93$ & $+51^\circ44'51\farcs96$ &
0.049 & Oct2001-Jan2002 & 52197-52288 & 70.2  & 22.4 \\

\multicolumn{1}{l}{\bf 1ES0229+200}  & $02^h32^m48\fs46$ & $+20^\circ17'16\farcs20$ &
0.139 & Oct2001-Jan2002 & 52193-52289 & 78.3  & 17.0 \\

\multicolumn{1}{l}{\bf 1ES0806+524}  & $08^h09^m49\fs15$ & $+52^\circ18'58\farcs70$ &
0.138 & Nov2001-Mar2002 & 52228-52348 & 69.6  & 23.5 \\

\multicolumn{1}{l}{\bf RGBJ1117+202} & $11^h17^m06\fs20$ & $+20^\circ14'07\farcs00$ &
0.139 & Dec2001-Feb2002 & 52265-52317 & 78.3  & 18.3 \\

\multicolumn{1}{l}{\bf 1ES1426+428}  & $14^h28^m32\fs66$ & $+42^\circ40'20\farcs60$ &
0.129 & Feb2001-Jun2001 & 51940-52073 & 79.3  & 19.0 \\

\multicolumn{1}{l}{\bf 1ES1553+113}  & $15^h55^m43\fs04$ & $+11^\circ11'24\farcs38$ &
0.360 & Apr2002-May2002 & 52373-52407 & 69.2  & 23.0 \\

\multicolumn{1}{l}{\bf RGBJ1725+118} & $17^h25^m04\fs36$ & $+11^\circ52'15\farcs20$ &
0.018 & Apr2002-Jul2002 & 52374-52460 & 69.9  & 22.0 \\

\multicolumn{1}{l}{\bf 1ES1959+650} & $19^h59^m59\fs85$ & $+65^\circ08'54\farcs67$ & 0.048 &
May2002-Jul2002 & 52410-52463 & 56.8 & 40.1 \\ 
\enddata 
\tablecomments{Max Ele corresponds to the maximum elevation that 
the object reaches as observed from the Whipple observatory (latitude=$31^\circ$ 57.6' N).}
\tablenotetext{(a)}{Tentative redshift.}
\tablenotetext{(*)}{An observing season spans from September to July. Because of
extreme weather conditions in Southern Arizona no observations are taken
during August.}
\end{deluxetable}

\clearpage

\begin{deluxetable}{llc}
\tablecaption{Image parameter cuts used in this work (Supercuts 2000). \label{SC2000}}
\tablewidth{0pt}
\tablehead{
\multicolumn{2}{l}{Parameter}    & Cut  \\
}
\startdata
\multicolumn{3}{l}{{\bf Cleaning$^{(a)}$}}\\
\tableline
& Picture      & 4.25$\sigma$ \\
& Boundary     & 2.25$\sigma$ \\
\tableline
\multicolumn{3}{l}{{\bf Image}}\\
\tableline
& Length             & $0.13~^\circ$ - $0.25~^\circ$ \\ 
& Width              & $0.05~^\circ$ - $0.12~^\circ$ \\
& Distance           & $0.4~^\circ$ - $1.0~^\circ$   \\
& Alpha              & $\le 15~^\circ$               \\
& Length/Size$^{(b)}$& $<0.0004~^\circ$/d.c.$^{(c)}$ \\
& Max1$^{(d)}$       & $>$ 30.~d.c.$^{(c)}$          \\
& Max2$^{(d)}$       & $>$ 30.~d.c.$^{(c)}$          \\
& \multicolumn{2}{l}{Three neighboring pixels have to be above the}\\
& \multicolumn{2}{l}{picture cut.}\\
\enddata
\tablenotetext{(a)}{ Only pixels with a significant signal above the background
are considered as part of the image and thus used in the image parameterisation. 
The signals in the rest of the pixels are set to zero and ignored in the
analysis. This process is know as image cleaning. See \cite{Reynolds93} for details.}
\tablenotetext{(b)}{ Size refers to the sum of the intensities over the 
pixels that constitute the image, in d.c.}
\tablenotetext{(c)}{ d.c. = digital counts}
\tablenotetext{(d)}{ Max1 and Max2 refer to the highest and second highest
pixel intensity in the image in d.c.}
\end{deluxetable}

\clearpage

\begin{deluxetable}{lccccccccccc}
\tabletypesize{\scriptsize} 
\tablecaption{Summary and results of observations
of the 8 X-ray selected BL Lacs considered in this work. Data have been analysed using Supercuts
2000 (Table \ref{SC2000}). $\langle \Theta \rangle$ refers to the mean zenith
angle of the observations. Flux upper limits (f.u.) are given in units of
$10^{-11}$~erg~cm$^{-2}$~s$^{-1}$.
\label{tev_observations}}
\tablewidth{0pt}
\tablehead{
 & Mode & Exp. & $\langle \Theta \rangle$ & Events & Events$^{(a)}$  &
 S & Rate & \multicolumn{2}{c}{U.L$_{>0.39\mathrm{TeV}}^{97\% \mathrm{c.l.}}$}        &
  \multicolumn{2}{c}{U.L$_{>1\mathrm{TeV}}^{97\% \mathrm{c.l.}}$} \\

Source &      &   (h) &$(^\circ)$& (RAW)  &($\alpha \le 15^\circ$) & ($\sigma$)
 &($\gamma$/min) &  (c.u.) & (f.u.) & (c.u.) & (f.u.)\\
}
\startdata
{\bf 1ES0033+595}& ON   & 12.02 & 29   &  967755  & 3794 & 0.42
              &  0.05 $\pm$ 0.12 & 0.11 & 0.97 & 0.03 & 0.24 \\
              & OFF &                12.04 & 26   &  997066  & 4023 (0.93)$^{(b)}$&
              & (5.56 $\pm$ 0.09)$^{(c)}$ &      &      & & \\
\tableline
{\bf 1ES0120+340}& ON   & 5.05 &  8.0 & 436321  & 1241 &-0.49
              & -0.08 $\pm$ 0.16 & 0.12 & 1.00 & 0.03 & 0.25 \\
              & OFF &                 7.44 & 10.1 & 647953  & 2034 (0.62)$^{(b)}$&
              & (4.55 $\pm$ 0.10)$^{(c)}$ &      &      & & \\
\tableline
{\bf RGBJ0214+517}& ON  & 6.05  & 22.4 & 557988 & 1923 & 0.64
              &  0.11 $\pm$ 0.17 & 0.17 & 1.51 & 0.04 & 0.37 \\
              & OFF &                 6.05  & 22.3 & 556078 & 1860 (1.01)$^{(b)}$&
              & (5.12 $\pm$ 0.12)$^{(c)}$ &      &      & & \\
\tableline
{\bf 1ES0229+200}& ON   & 14.69 & 17.0     &  1341365 & 4570 & 0.74
              &  0.08 $\pm$ 0.11 & 0.11 & 0.97 & 0.03 & 0.24 \\
              & OFF &                 15.82 & 15.0     &  1439227 & 4727 (0.95)$^{(b)}$&
              & (4.98 $\pm$ 0.07)$^{(c)}$ &      &      & & \\
\tableline
{\bf 1ES0806+524}& ON   & 18.70 & 23.5  &1856253 & 4956 & 0.14
              &  0.01 $\pm$ 0.09 & 0.08 & 0.67 & 0.02 & 0.16 \\
              & OFF &                 18.83 & 23.0  &1867487 & 4942 (1.00)$^{(b)}$&
              & (4.37 $\pm$ 0.06)$^{(c)}$ &      &      & & \\
\tableline
{\bf RGBJ1117+202}& ON  & 3.26 &  18.3    &  348436 &  933 & 0.47
              &  0.10 $\pm$ 0.22 & 0.21 & 1.84 & 0.05 & 0.45 \\
              & OFF &                 3.26 &  19.0    &  350701 &  915 (0.99)$^{(b)}$&
              & (4.69 $\pm$ 0.15)$^{(c)}$ &      &      & & \\
\tableline
{\bf 1ES1553+113}& ON  & 2.82 & 23.0     & 238032 & 554 & 0.77
              &  0.15 $\pm$ 0.20 & 0.19 & 1.62 & 0.05 & 0.40 \\
              & OFF &                 3.72 &  22.3     & 326616 & 743 (0.71)$^{(b)}$&
              & (3.32 $\pm$ 0.12)$^{(c)}$ &      &      & & \\
\tableline
{\bf RGBJ1725+118}& ON  & 2.33 &  22.2   & 187298 &  489 & 0.67
              &  0.15 $\pm$ 0.22 & 0.23 & 1.98 & 0.06 & 0.49 \\
              & OFF &                 2.33 & 22.2   & 189671 &  467 (1.00)$^{(b)}$&
              & (3.34 $\pm$ 0.16)$^{(c)}$ &      &      & & \\
\enddata
\tablenotetext{(a)}{ Events after image cuts (Supercuts 2000).}
\tablenotetext{(b)}{ Scaling factor (SCF).}
\tablenotetext{(c)}{ Background rate in counts/min.}
\end{deluxetable}

\clearpage

\begin{deluxetable}{lcccccc}
\tabletypesize{\scriptsize} \tablecaption{Comparison between predicted and
measured TeV fluxes for the BL Lac sample considered in this work. Whipple
values correspond to the ones obtained in this work. Predicted flux values
have been converted into Crab units using the integral Crab flux,
F$_{\mathrm{Crab}}(>$ 0.3~TeV) = 12.27 $\times$
$10^{-11}$~erg~cm$^{-2}$~s$^{-1}$. The predicted fluxes are given by the
modified version of the parameterisation of the SED according to
\cite{Fossati98} / according to the SSC model both described in
\cite{Costamante01}. Also predicted flux values are given according to
\cite{Stecker96} where available.\label{tev_comparisson} }
\tablewidth{0pt} 
\tablehead{ 
& \multicolumn{2}{c}{\bf Flux Prediction} & \multicolumn{1}{c}{IR$^{(a)}$} &
{\bf Whipple} & {\bf CAT$^{(3)}$} & {\bf HEGRA$^{(4)}$}\\

&  \multicolumn{2}{c}{F($>$0.3~TeV)} &  & F($>$0.39~TeV) & F($>$0.25~TeV) & \\ 

& FOSS/SSC$^{(1)}$ & Stecker$^{(2)}$ &  & U.L.$^{97\% \mathrm{c.l.}}$ & U.L.$^{99.7\% \mathrm{c.l.}}$ &
U.L.$^{99.9\% \mathrm{c.l.}}$\\

Source&   (c.u.)   &   (c.u.)  & $(\%)$ & (c.u.)         &  (c.u.)           &   (c.u.) \\  
}
\startdata
\multicolumn{1}{l}{\bf 1ES0033+595}  & 0.17/0.021 &
\multicolumn{1}{c}{.....} & 46 & 0.11 0.16$^{(b)}$ {\bf 0.23}$^{(c)}$
& 0.45  & ..... \\ 

\multicolumn{1}{l}{\bf 1ES0120+340}  & 0.02/0.025 &
\multicolumn{1}{c}{.....} & 88 & 0.12 0.18$^{(b)}$ {\bf 0.34}$^{(c)}$
& ..... & 0.03$_{>0.75~\mathrm{TeV}}$ \\ 

\multicolumn{1}{l}{\bf RGBJ0214+517} & 0.48/0.006 &
\multicolumn{1}{c}{.....} & 30 & 0.17 0.25$^{(b)}$ {\bf 0.32}$^{(c)}$
& 0.39  & ..... \\ 

\multicolumn{1}{l}{\bf 1ES0229+200}  & 0.08/0.026 & 
\multicolumn{1}{c}{0.02}  & 63 & 0.11 0.16$^{(b)}$ {\bf 0.26}$^{(c)}$
& ..... & 0.25$_{>0.54~\mathrm{TeV}}$ \\ 

\multicolumn{1}{l}{\bf 1ES0806+524}  & 0.11/..... &
\multicolumn{1}{c}{.....} & 63 & 0.08 0.12$^{(b)}$ {\bf 0.19}$^{(c)}$
& 0.67  & ..... \\   

\multicolumn{1}{l}{\bf RGBJ1117+202} & 0.09/0.008 & 
\multicolumn{1}{c}{.....} & 63 & 0.21 0.31$^{(b)}$ {\bf 0.50}$^{(c)}$
& ..... & ..... \\

\multicolumn{1}{l}{\bf 1ES1553+113}  & 0.02/0.035 & 
\multicolumn{1}{c}{.....} & 95 & 0.19 0.28$^{(b)}$ {\bf 0.55}$^{(c)}$
& ..... & ..... \\ 

\multicolumn{1}{l}{\bf RGBJ1725+118} & 1.04/0.001 & 
\multicolumn{1}{c}{.....} & 12 & 0.23 0.34$^{(b)}$ {\bf 0.38}$^{(c)}$
& ..... & 0.09$_{>0.69~\mathrm{TeV}}$ \\  
\enddata
\tablenotetext{(a)}{Using the optical depth given in
\cite{deJagger02}. Estimated absorption of the gamma-ray photon flux 
between 300~GeV and 10~TeV assuming a Crab-like spectrum.}
\tablenotetext{(b)}{U.L.$^{97\% \mathrm{c.l.}}$ above 300~GeV. The correction has been
done by assuming a Crab-like integral spectrum. Upper limits are increased by
$48\%$ when going from 390~GeV to 300~GeV.}
\tablenotetext{(c)}{U.L.$^{97\% \mathrm{c.l.}}$ above 300~GeV IR corrected.}
\tablerefs{(1) Costamante $\&$ Ghisselini 2002; (2) Stecker et al. 1996;
(3) Piron 2000; (4) Tluczykont et al. 2001, Poster Session OG198 27th ICRC,
Hamburg and Aharonian et al. 2000.}
\end{deluxetable}

\clearpage

\begin{deluxetable}{lcccccccc}
\tabletypesize{\scriptsize} 
\tablecaption{Mean RXTE (ASM) X-ray fluxes (where available) during various time
periods for the sources in this work. The mean flux in each case is the 
weighted mean for the appropiate interval, and is expressed in mCrab units. 
\label{xray_meanfluxes_tab} }
\tablewidth{0pt}
\tablehead{
Name$^{(a)}$   & 1997-2002$^{(b)}$ & 1997 & 1998 & 1999 & 2000 & 2001 & 2002$^{(b)}$ &
\multicolumn{1}{c}{$\gamma_{Obs.}^{(c)}$}\\
& mCrab & mCrab & mCrab & mCrab & mCrab & mCrab & mCrab & mCrab \\
}
\startdata
0120   & 2.15 $\pm$ 0.11 & ..... & 2.52 $\pm$ 0.27 &
2.74 $\pm$ 0.22 & 1.25 $\pm$ 0.23 & 2.35 $\pm$ 0.25 &
1.63 $\pm$ 0.36 & 2.33 $\pm$ 0.71 \\
0214 & 1.66 $\pm$ 0.08 & 1.72 $\pm$ 0.14 & 1.54 $\pm$ 0.17 &
1.61 $\pm$ 0.21 & 1.78 $\pm$ 0.20 & 1.70 $\pm$ 0.22 & 1.49 $\pm$ 0.27 &
0.89 $\pm$ 0.37 \\
0229   & 1.78 $\pm$ 0.08 & 2.64 $\pm$ 0.16 & 1.20 $\pm$ 0.17  & 1.11
$\pm$ 0.22 & 1.70 $\pm$ 0.20 & 1.86 $\pm$ 0.24 & 2.10 $\pm$ 0.31 & 1.05 $\pm$
0.37 \\
0806   & 1.51 $\pm$ 0.07 & 1.07 $\pm$ 0.15 & 1.57 $\pm$ 0.15 &
1.42 $\pm$ 0.18 & 1.48 $\pm$ 0.18 & 1.85  $\pm$ 0.19 & 2.19 $\pm$ 0.24 &
1.84 $\pm$ 0.29 \\
{\bf 421}   & 8.68 $\pm$ 0.07 & 6.20 $\pm$ 0.16  & 12.52 $\pm$ 0.19 & 4.83
$\pm$ 0.19 & 12.60 $\pm$ 0.19 & 14.03 $\pm$ 0.23 & 7.69 $\pm$ 0.24 &
19.88 $\pm$ 0.29$^{(d)}$ \\

{\bf 1426}   & 2.88 $\pm$ 0.08   & 2.30 $\pm$ 0.18 & 2.99 $\pm$ 0.17 & 2.68
$\pm$ 0.19 & 3.22 $\pm$ 0.19 & 3.42 $\pm$ 0.20 & 2.75 $\pm$ 0.22 &
3.69 $\pm$ 0.33 \\

{\bf 501}   & 7.32 $\pm$ 0.06 & 15.57 $\pm$ 0.16 & 7.65 $\pm$ 0.15 & 5.69
$\pm$ 0.16 & 5.39 $\pm$ 0.17 & 4.26 $\pm$ 0.17 & 4.34 $\pm$ 0.21 &
17.35 $\pm$ 0.27$^{(e)}$ \\

1725  & 2.64 $\pm$ 0.09 & 2.65 $\pm$ 0.20 & 2.40 $\pm$ 0.20 & 2.54
$\pm$ 0.22 & 1.98 $\pm$ 0.23 & 3.16 $\pm$ 0.22 & 3.27 $\pm$ 0.30 & 2.56 $\pm$
0.44 \\
{\bf 1959}  & 3.75 $\pm$ 0.06 & 2.57 $\pm$ 0.12 & 2.62 $\pm$ 0.14 & 4.06
$\pm$ 0.14 & 5.05 $\pm$ 0.16 & 4.54 $\pm$ 0.15 & 4.77 $\pm$ 0.19 & 6.48 $\pm$
0.39 \\
\enddata
\tablenotetext{(a)} {0120: 1ES0120+340, 0214: RGBJ0214+517, 0229: 1ES0229+200,
0806: 1ES0806+524, 421: Mrk421, 1426: 1ES1426+428,\\
501: Mrk501, 1725: RGBJ1725+118, 1959: 1ES1959+650.}
\tablenotetext{(b)} {Extends only up to 2002 July.}
\tablenotetext{(c)} {Refers to the period of time in which TeV observations where taken.}
\tablenotetext{(d)} {Period from 2000 November to 2001 May.}
\tablenotetext{(e)} {Period from 1997 March to 1997 June.}
\end{deluxetable}




\end{document}